\documentclass[11pt,letterpaper]{article}
\usepackage{epsfig}
\usepackage[]{natbib}
\usepackage{amsmath,amsthm}
\usepackage{amssymb}
\usepackage{graphicx,color,caption}
\usepackage{nth}
\usepackage{microtype}
\usepackage{hyperref} 
\usepackage[active]{srcltx}

\newcommand*{\defeq}{\mathrel{\vcenter{\baselineskip0.5ex\lineskiplimit0pt\hbox{\scriptsize.}\hbox{\scriptsize.}}}=}

\newcommand{\argmin}{\mathop{\mathrm{arg min}}}

\newcommand{\im}{{\sf Im}}

\newcommand{\bone}{{\bf 1}}
\newcommand{\C}{{\mathbb{C}}}
\newcommand{\R}{{\mathbb{R}}}

\newcommand{\be}{\begin{equation}}
\newcommand{\ee}{\end{equation}}

\theoremstyle{plain}

\def\scale{1.0}

\allowdisplaybreaks

\begin{document}

\title{Numerical Implementation of the QuEST Function\thanks{The
    authors wish to thank Edgar Dobriban (Stanford University),
    Jonathan Fletcher (University of Strathclyde), Matan Gavish
    (Stanford University), Na Huang (London School of Economics and
    Political Science), Wen
    Jun (National University of Singapore), Tatsuya Kubokawa
    (University of Tokyo), Clifford Lam (London School of Economics
    and Political Science),
    Artyom Lepold (University of Kaiserslautern), Stefano Monni
    (American University of Beirut), Nestor Parolya (University of
    Hannover), Simon Welsing (Technische \mbox{Universit\"at} M\"unchen), and Zhao Zhao (Huazhong University of Science and Technology) for testing early versions of this code. Any errors are ours.}}

\bigskip 

\author
 {$\,$ \\
 Olivier Ledoit\\
 Department of Economics \\
 University of Zurich\\
 CH-8032 Zurich, Switzerland\\
 {\tt olivier.ledoit@econ.uzh.ch}\\
\and
 $\,$ \\
 Michael Wolf\\
 Department of Economics \\
 University of Zurich\\
 CH-8032 Zurich, Switzerland\\
{\tt michael.wolf@econ.uzh.ch}}

\date
{January 2016}

\maketitle

\begin{abstract}
This paper deals with certain estimation problems involving the
covariance matrix in large dimensions. Due to the breakdown of
finite-dimensional asymptotic theory when the dimension is not negligible
with respect to the sample size, it is necessary to resort to an
alternative framework known as large-dimensional
asymptotics. Recently, \cite{ledoit:wolf:2015} have proposed an estimator
of the eigenvalues of the population covariance matrix that is
consistent according to a mean-square criterion under
large-dimensional asymptotics. It requires numerical inversion of a
multivariate nonrandom function which they call the QuEST
function. The present paper explains how to numerically implement the
QuEST function in practice through a series of six successive
steps. It also provides an algorithm to compute the Jacobian
analytically, which is necessary for numerical inversion by a
nonlinear optimizer. Monte Carlo simulations document the
effectiveness of the code.
\end{abstract}

\begin{tabbing}
            \> random matrix theory, spectrum estimation.
\end{tabbing}
\noindent 
JEL CLASSIFICATION NOS: C13, C61, C87.

\newpage

\section{Introduction}

Many data sets in econometrics, biostatistics and electrical
engineering, among a host of other fields, contain large numbers of
related variables. The estimation of the covariance matrix poses
challenging statistical problems when the dimension is not small
relative to sample size. Approximations that are valid under
traditional asymptotics, that is, when the dimension remains fixed
while the sample size goes to infinity, perform poorly. This is why
attention has turned to {\em large-dimensional asymptotics} where  the
dimension and  the sample size go to infinity together, with their
ratio converging to a finite, nonzero limit called the {\it
  concentration (ratio)}. 

Under large-dimensional asymptotics, the sample eigenvalues are not
consistent estimators of the population eigenvalues. A new estimator
for the population eigenvalues under large-dimensional asymptotics was
recently introduced by \cite{ledoit:wolf:2015}. It hinges critically
on a multivariate nonrandom function called the QuEST function. This
acronym stands for {\it Quantized Eigenvalues Sampling
  Transform}. \cite{ledoit:wolf:2015} provide the mathematical
definition of the QuEST function, but do not provide any details about
numerical implementation. The problem of numerical implementation is
non-trivial, due to the complexity of the definition of the QuEST
function. A direct application of this method is the optimal
estimation of the covariance matrix in the class of
rotation-equivariant estimators introduced by \cite{stein:1975,stein:1986}
under various loss functions; see \cite{ledoit:wolf:stein:2013}.

This paper explains how to numerically implement the QuEST function accurately and efficiently. In addition, given that the estimation of the population eigenvalues requires numerically inverting the QuEST function using a nonlinear optimizer, we also give the Jacobian analytically. 

Section \ref{sec:litrev} reviews the literature on this
subject. Section \ref{sec:def} gives the definition of the problem
that will be solved numerically. Sections
\ref{sec:support}--\ref{sec:final} describe in detail the six
steps needed to implement the QuEST function numerically, delineating
all the mathematical results that are needed along the way. Section
\ref{sec:mc} provides extensive Monte Carlo simulations. \mbox{Section~\ref{sec:conclusion} concludes.}

\section{Literature Review}
\label{sec:litrev}

\subsection{Estimation of the Population Covariance Matrix Eigenvalues}

\cite{karoui:2008} proposed a way to estimate the empirical c.d.f.\ of
population eigenvalues under large-dimensional asymptotics using a
different approach than the QuEST function. However, the code
executing his algorithm was not made available to other researchers in
the field, and those who tried to replicate it themselves did not
enjoy much success. The state of affairs is aptly summarized by
\cite{li2013estimation}:
\begin{quote}
 Actually, the general
approach in \cite{karoui:2008} has several implementation issues that seem to be responsible for its relatively low
performance as attested by the very simple nature of provided simulation results. 
\end{quote}

There are three reasons why the same criticisms cannot be levelled
against the \mbox{QuEST~function}: first, a Matlab executable implementing the
QuEST function has already been used independently by
\cite{welsing:2015}, \cite{ito:kubokawa:2015},
\cite{na:fryzlewicz:2015}, and \cite{lam:2016}, among
others\footnote{The Matlab executable can be downloaded at
{\href{http://www.econ.uzh.ch/en/people/faculty/wolf/publications.html}
{http://www.econ.uzh.ch/en/people/faculty/wolf/publications.html}}
under the link ``Programming Code''.};
second, the present paper
opens up the code of the QuEST function and its Jacobian to the
general public for inspection and potential improvements; and third, Section
\ref{sec:mc} provides an extensive Monte Carlo study with nearly a
third of a million simulations across a variety of challenging
scenarios.

Apart from \cite{karoui:2008}, other proposals have been put forward, making this field one of the most active ones in multivariate analysis in recent years. 
\begin{itemize}
\item \cite{rao2008statistical} provide a solution when the population
  spectrum has a staircase structure, typically with half of the
  eigenvalues equal to one and the rest equal to two. The ability of
  this approach to handle the general case where there can be up to
  $p$ distinct population eigenvalues, with $p$ going to infinity, is not
  established. 
\item \cite{mestre:2008b} provides a solution when the concentration
  ratio $c=p/n$ is sufficiently small and/or the distinct population
  eigenvalues sufficiently far from one another, that is, when the  sample
  eigenvalues display what is known as ``spectral separation''. This
  is a favorable situation where the sample eigenvalues are grouped into
  easily identifiable clusters, each cluster corresponding to one
  single population eigenvalue (which can have multiplicity higher
  than one). Monte Carlo simulations assume no more than four distinct
  population eigenvalues.

\item \cite{bai2010estimation} propose a solution  based on the method
  of moments when the parametric dimension of the population spectrum
  is finite. They demonstrate good behavior up to order four. 
\item \cite{chen2011model} elaborate on the previous paper by providing more rigorous justification of the method when the model order is unknown. But Monte Carlo simulations only go to order three.
\item \cite{yao2012eigenvalue} can be seen as a cross between the papers of \cite{mestre:2008b} and \cite{bai2010estimation}, but also requiring a finite number of distinct population eigenvalues. In practice, Monte Carlo simulations provided by the authors do not go above three distinct population eigenvalues.
\end{itemize}
The common point between all these other methods is that they do not purport to address the general case. They work with a finite number of degrees of freedom (in practice no more than four) in the choice of the population spectral distribution, whereas the real number is $p$, which goes to infinity. This is why it is important to avoid the criticisms that have been levelled at the only other ostensibly general approach, that of \cite{karoui:2008}, by fully explaining how to numerically implement the QuEST function, and by providing extensive Monte Carlo simulations showing that it works in practice under a wide variety of circumstances. 

Finally, we should note that \cite{dobriban2015efficient} also provides a numerical method for solving the \cite{marcenko:pastur:1967} equation. He does not compute the QuEST function explicitly, and does not provide the Jacobian analytically. As a result, numerical inversion is very difficult, but his paper is not focused on the problem of recovering the population eigenvalues.

\subsection{Potential Applications}

The numerical implementation of the QuEST function given in this paper is essential for the estimation of the population eigenvalues, which in turn is essential for computing the optimal nonlinear shrinkage of the covariance matrix under large-dimensional asymptotics. Many fields are interested in shrinking the covariance matrix when the number of variables is high:
\begin{description}
\item[Acoustics] Optimally removing noise from signals captured from an array of hydrophones \citep{zhang2009robust}.
\item[Cancer Research] Mapping out the influence of the Human Papillomavirus (HPV) on gene expression \citep{pyeon2007fundamental}.
\item[Chemistry] Estimating the temporal autocorrelation function (TACF) for fluorescence correlation spectroscopy \citep{guo2012bayesian}.
\item[Civil Engineering] Detecting and identifying vibration--based bridge damage through Random Coefficient Pooled (RCP) models \citep{michaelides2011vibration}. 
\item[Climatology] Detecting trends in average global temperature through the optimal fingerprinting method \citep{ribes2013application}.
\item[Econometrics] Specifying the target covariance matrix in the Dynamic Conditional Correlation (DCC) model to capture time-series effects in the second moments \citep{hafner2012estimation}.
\item[Electromagnetics] Studying correlation between reverberation chamber measurements collected at different stirrer positions \citep{pirkl2012reverberation}
\item[Entertainment Technology] Designing a video game controlled by performing tricks on a skateboard \citep{anlauff2010method}.
\item[Finance] Reducing the risk in large portfolios of stocks \citep{jagannathan:ma:2003}.
\item[Genetics] Inferring large-scale covariance matrices from functional genomic data \citep{schaefer:strimmer:2005}.
\item[Geology] Modeling multiphase flow in subsurface petroleum reservoirs with the iterative sto\-chastic ensemble method (ISEM) on inverse problems \citep{elsheikh2013iterative}.
\item[Image Recognition] Detecting anomalous pixels in hyperspectral imagery \citep{bachega2011evaluating}.
\item[Neuroscience] Calibrating brain-computer interfaces \citep{lotte2009efficient}.
\item[Psychology] Modeling co-morbidity patterns among mental disorders \citep{markon2010modeling}.
\item[Road Safety Research] Developing an emergency braking assistance system \citep{haufe2011eeg}.
\item[Signal Processing] Combining data recorded by an array of sensors to minimize the noise \citep{chen2010shrinkage}.
\item[Speech Recognition] Automatically transcribing records of phone conversations \citep{bell2009diagonal}.
\end{description} 
Up until now, these fields have had to satisfy themselves with {\em
  linear} shrinkage estimation of the covariance matrix
\citep{ledoit:wolf:2003,ledoit:wolf:2004a}. However this approach is asymptotically suboptimal in the class of rotation-equivariant estimators relative to nonlinear shrinkage, which requires numerical implementation of the QuEST function. The present paper makes this new and improved method universally available in practice.

\section{Definition of the QuEST Function}
\label{sec:def}

The mathematical definition of the QuEST function is given by \cite{ledoit:wolf:2015}. It~is reproduced here for convenience. For any positive integers $n$ and $p$,
the QuEST function, denoted by $Q_{n,p}$, is the nonrandom multivariate function given by:
\begin{align}
Q_{n,p}:[0,\infty)^p&\longrightarrow[0,\infty)^p \label{eq:population}\\
\mathbf{t}\defeq\left(t_1,\dots,t_p\right)^\prime&\longmapsto
Q_{n,p}(\mathbf{t})\defeq\left(q_{n,p}^1(\mathbf{t}),\ldots,q_{n,p}^p(\mathbf{t})\right)^\prime
,\label{eq:quest2}
\end{align}
where
\begin{align}
\forall i=1,\ldots,p\qquad 
q_{n,p}^i(\mathbf{t})&\defeq
p\displaystyle\int_{(i-1)/p}^{i/p}\left(F_{n,p}^\mathbf{t}\right)^{-1}(v)\,dv~,\label{eq:sample}\\
\forall v\in[0,1]\qquad \left(F_{n,p}^\mathbf{t}\right)^{-1}(v)&\defeq \sup\{x\in\R:F_{n,p}^\mathbf{t}(x)\leq v\}~,\label{eq:inverse}\\
\forall x\in\R\qquad F_{n,p}^\mathbf{t}(x) &\defeq 
\begin{cases}
\displaystyle\max\left(1-\frac{n}{p},\frac{1}{p}\sum_{i=1}^p\bone_{\{t_i=0\}} \right)&\text{if $x=0$~,}\\
\displaystyle\lim_{\eta\to0^+}\frac{1}{\pi}\int_{-\infty}^x
\im\left[m_{n,p}^\mathbf{t}(\xi+i\eta)\right]\,d\xi & \text{otherwise~,}
\end{cases}\label{eq:questF},
\end{align}
and $\forall z\in\C^+\quad m\defeq m_{n,p}^\mathbf{t}(z)$ is the unique solution in the set 
\be
\left\{m\in\C:-\frac{n-p}{nz}+\frac{p}{n}\,m\in\C^+\right\}\label{eq:questset}
\ee
to the equation 
\be
m=\frac{1}{p}\sum_{i=1}^p\frac{1}{\displaystyle t_i\left(1-\frac{p}{n}-\frac{p}{n}\,z\,m\right)-z}~.\label{eq:questMP}
\ee
The QuEST function is a natural discretization of equation (1.4) of
\cite{silverstein:1995}, which is itself a reformulation of equation
(1.14) of \cite{marcenko:pastur:1967}. The basic idea is that $p$
represents the matrix dimension, $n$ the sample size,
$\mathbf{t}\defeq\left(t_1,\dots,t_p\right)^\prime$ the population
eigenvalues,
$Q_{n,p}(\mathbf{t})\defeq\left(q_{n,p}^1(\mathbf{t}),\ldots,q_{n,p}^p(\mathbf{t})\right)^\prime$
the sample eigenvalues, $F_{n,p}^\mathbf{t}$ the limiting empirical
c.d.f\ of sample eigenvalues, and $m_{n,p}^\mathbf{t}$ its
\cite{stieltjes:1894} transform. A fundamental result in
large-dimensional asymptotics is that the relationship between the
population spectral distribution and the sample spectral distribution
is nonrandom in the limit. Figure \ref{fig:MParea}, publicized by
Jianfeng \cite{yao:2015} in a conference presentation, gives a heuristic view of the area where Mar\v{c}enko-Pastur asymptotic theory is more useful (labelled ``MP area'') vs.\ the area where standard fixed-dimension asymptotic theory applies (labelled ``Low-dim area'').
\begin{center}
\captionsetup{type=figure}
\includegraphics[scale=0.4,trim={1.4cm 1.3cm 5cm 2.7cm},clip=true]{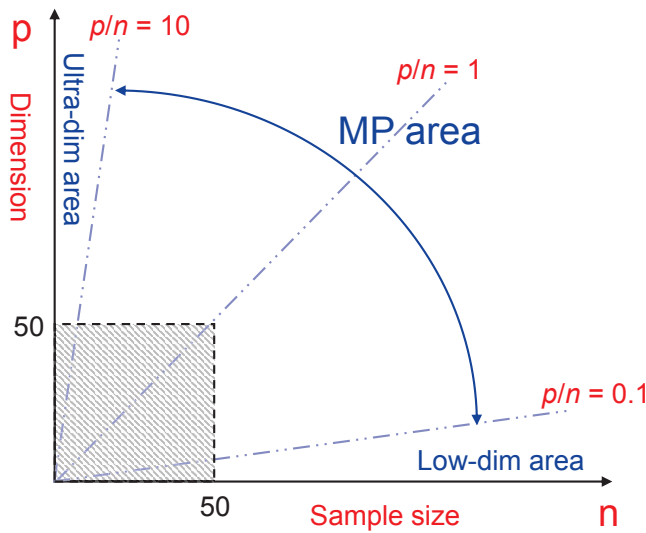}
\captionof{figure}{Heuristic comparison of the area of relevance of Mar\v{c}enko-Pastur asymptotics vs.\ traditional fixed-dimension asymptotics.}\label{fig:MParea}
\end{center}
This insight is further developed in the recent book by \cite{yao2015large}. Readers interested in the background from probability theory may also consult the authoritative monograph by \cite{bai:silverstein:2010}.

The importance of the QuEST function is twofold. First, inverting it
numerically yields an estimator of the population eigenvalues that is
consistent under large-dimensional asymptotics. Second, once this has
been achieved, it is possible to use Theorem 2 of
\cite{ledoit:peche:2011} to construct shrinkage estimators of the
covariance matrix that are asymptotically optimal with respect to a
given loss function in the $p$-dimensional space of
rotation-equivariant estimators introduced by \cite{stein:1975,stein:1986}. \cite{ledoit:wolf:stein:2013} derive
the optimal shrinkage formula for five different loss functions, and
\cite{ledoit:wolf:goldi:2014} for a sixth.

Numerical implementation of the QuEST function consists in a series of six successive operations: 1) finding the support of $F_{n,p}^\mathbf{t}$; 2) choosing a grid that covers the support; 3)~solving equation \eqref{eq:questMP} on the grid; 4) computing the sample spectral density; 5) integrating it to obtain the empirical c.d.f.~of sample eigenvalues; and 6) interpolating the c.d.f~to compute sample eigenvalues as per equation (\ref{eq:sample}). Each of these steps is detailed below.

\section{Support}
\label{sec:support}

In what follows we omit the subscripts and superscript of $F_{n,p}^\mathbf{t}$ in order to simplify the notation. 
We do not work directly with $F$ but with $u$, which is defined by:
\begin{align*}
u & \defeq u(z) \defeq -\frac{1}{m_{\underline{F}}(z)}\\
m_{\underline{F}}(z) &\defeq \frac{c-1}{z}+c\,m_F(z)\\
m_{F}(z) &\defeq \int_{-\infty}^{+\infty}\frac{1}{\lambda-z}dF(\lambda)~.
\end{align*}
There is a direct mapping between $F$-space and $u$-space, as explained in Section 2 of \cite{ledoit:wolf:2012}. Numerically it is more judicious to work in $u$-space.

To determine the image of the support of $F$ in $u$-space, we first
need to group together the population eigenvalues
$\tau_1,\ldots,\tau_p$ that are equal to one another and, if
necessary, discard those that are equal to zero. Let us say that there
are $K$ distinct nonzero population eigenvalues $0<t_1<\ldots<t_K$. We
can associate them with their respective weights: if $j$ elements of
the vector $(\tau_1,\ldots,\tau_p)$ are equal to $t_k$ then the
corresponding weight is $w_k \defeq j/p$.

\subsection{Spectral Separation}

Now we look for spectral separation between $t_{k}$ and $t_{k+1}$ ($k=1,\ldots,K-1$). This is done in two stages. First we run a quick test to see whether we can rule out spectral separation {\em a~priori}. Second, if the test is inconclusive, we do the full analysis to ascertain whether spectral separation does indeed occur.

\subsubsection{Necessary Condition}

Spectral separation occurs between $t_{k}$ and $t_{k+1}$ if and only if
$$\exists u\in(t_{k},t_{k+1}),\;\exists v\in(0,+\infty)\quad\mbox{s.t.}\quad
\im\left[u-cu\sum_{k=1}^K\frac{w_kt_k}{t_k-(u+iv)}\right]=0~,$$
which is equivalent to
\be
\exists u\in(t_{k},t_{k+1})\quad\mbox{s.t.}\quad\sum_{j=1}^K\frac{w_j t_j^2}{(t_j-u)^2}<\frac{1}{c}~.\label{eq:separation}
\ee 
Equation (\ref{eq:separation}) is equivalent to the function $x_F(m)$ defined in equation (1.6) of \cite{silverstein:choi:1995} being strictly increasing at $m=-1/u$. Section 4 of \cite{silverstein:choi:1995} explains how this enables us to determine the support.

Call $\varphi(u)$ the function on the left-hand side of equation (\ref{eq:separation}). We can decompose it into
\begin{align*}
\varphi(u)&=\theta_k(u)+\psi_k^L(u)+\psi_k^R(u)\\
\mbox{where}\qquad\theta_k(u) &\defeq \frac{w_kt_k^2}{(t_k-u)^2}+\frac{w_{k+1}t_{k+1}^2}{(t_{k+1}-u)^2}~,\\
\psi_k^L(u) &\defeq \sum_{j=1}^{k-1}\frac{w_j t_j^2}{(t_j-u)^2}~,\\
\mbox{and}\qquad\psi_k^R(u) &\defeq \sum_{j=k+2}^{K}\frac{w_j t_j^2}{(t_j-u)^2}~.
\end{align*}
It is easy to see that the function $\theta_k(\cdot)$ is convex over the interval $(t_{k},t_{k+1})$, diverges to $+\infty$ near $t_k$ and $t_{k+1}$, and attains its minimum at
\be
\widehat{x}_k\defeq(t_kt_{k+1})^{2/3}\;\frac{w_k^{1/3}\tau_{k+1}^{1/3}+w_{k+1}^{1/3}\tau_{k}^{1/3}}{w_k^{1/3}\tau_{k}^{2/3}+w_{k+1}^{1/3}\tau_{k+1}^{2/3}}~,\label{eq:xhat}
\ee
therefore a lower bound for $\theta_k(\cdot)$ on $(t_{k},t_{k+1})$ is $\theta_k(\widehat{x}_k)$.

It is also easy to see that the function  $\psi_k^L(\cdot)$ is decreasing over the interval $(t_{k},t_{k+1})$; therefore, it attains its minimum at $t_{k+1}$ and is bounded from below by $\psi_k^L(t_{k+1})$. Conversely, the function  $\psi_k^R(\cdot)$ is increasing over the interval $(t_{k},t_{k+1})$, attains its minimum at $t_{k}$ and is bounded from below by $\psi_k^R(t_{k})$. Putting these three results together yields the following lower bound for $\varphi(\cdot)$:
$$\forall u\in(t_{k},t_{k+1})\qquad \varphi(u)\geq
\frac{w_kt_k^2}{(t_k-\widehat{x}_k)^2}
+\frac{w_{k+1}t_{k+1}^2}{(t_{k+1}-\widehat{x}_k)^2}
+\sum_{j=1}^{k-1}\frac{w_j t_j^2}{(t_j-t_{k+1})^2}
+\sum_{j=k+2}^{K}\frac{w_j t_j^2}{(t_j-t_k)^2}~,$$
where $\widehat{x}_k$ is given by equation (\ref{eq:xhat}).

Combining this bound with equation (\ref{eq:separation}) means that
\be
\frac{w_kt_k^2}{(t_k-\widehat{x}_k)^2}
+\frac{w_{k+1}t_{k+1}^2}{(t_{k+1}-\widehat{x}_k)^2}
+\sum_{j=1}^{k-1}\frac{w_j t_j^2}{(t_j-t_{k+1})^2}
+\sum_{j=k+2}^{K}\frac{w_j t_j^2}{(t_j-t_k)^2}<\frac{1}{c}\label{eq:necessary}
\ee
 is a necessary (but not sufficient) condition for spectral separation to occur between $t_{k}$ and~$t_{k+1}$. Thus, the numerical procedure can be made more efficient by first computing the quantity on the left-hand side of equation (\ref{eq:necessary}), comparing it to $1/c$, and discarding the interval $(t_{k},t_{k+1})$ in the case where it is higher than $1/c$. If, on the other hand, it is strictly lower than $1/c$, then further work is needed to ascertain whether spectral separation does indeed occur. In practice, checking this condition seems to save a lot of time by eliminating many intervals $(t_{k},t_{k+1})$, except perhaps when $c$ is very small and the population eigenvalues are very spread out.

\subsubsection{Necessary and Sufficient Condition}

Consider now some $k\in\{1,2,\ldots,K-1\}$ for which the condition in equation (\ref{eq:necessary}) does not hold. Given equation (\ref{eq:separation}), we need to find the minimum of $\varphi(\cdot)$ over $(t_{k},t_{k+1})$ and compare it to $1/c$. It is easy to check that $\varphi(\cdot)$ is strictly convex over $(t_{k},t_{k+1})$, therefore this minimum exists, is unique, and is the only zero in $(t_{k},t_{k+1})$ of the derivative function
$$\varphi'(u)=2\sum_{j=1}^K\frac{w_j t_j^2}{(t_j-u)^3}~.$$
Most numerical algorithms that find the zero of a function require as inputs two points $(\underline{x},\overline{x})$ such that the sign of the function is not the same at $\underline{x}$ as at $\overline{x}$. Finding two such points is the goal of the next step. There are three cases, depending on the sign of $\varphi'(\widehat{x}_k)$.
\begin{itemize}
\item $\varphi'(\widehat{x}_k)=0$:
Then the search is immediately over because $\varphi(\cdot)$ attains its minimum at $x^*_k\defeq\widehat{x}_k$. This would not happen generically unless $K=2$.
\item $\varphi'(\widehat{x}_k)<0$: In this case, given that
  $\varphi'(\cdot)$ is strictly increasing, the minimizer of
  $\varphi(\cdot)$ lies in the interval $(\widehat{x}_k,t_{k+1})$. We
  can feed the lower bound $\underline{x}=\widehat{x}_k$ into the
  numerical procedure that will find the zero of $\varphi'(\cdot)$. It
  would  be also tempting to set $\overline{x}\defeq t_{k+1}$, but
  unfortunately doing so would not be practicable because
  $\lim_{u\nearrow t_{k+1}}\varphi'(u)=+\infty$, and most numerical
  procedures perform poorly near singularity points. Therefore we need
  to find some $\overline{x}\in(\widehat{x}_k,t_{k+1})$ such that
  $\varphi'(\overline{x})>0$. Let $x^*_k$ denote the unique value in
  $(\widehat{x}_k,t_{k+1})$ such that $\varphi'(x^*_k)=0$. Then the
  fact that $w_{j}t_{j}^2/(t_{j}-u)^3$ is increasing in $u$ for any
  $j\in\{1,\ldots,K\}$ implies that the following inequalities hold:

\begin{align*}
\forall u\in(\widehat{x}_k,t_{k+1})\qquad\varphi'(u) &>
2\frac{w_kt_k^2}{(t_k-\widehat{x}_k)^3}
+2\frac{w_{k+1}t_{k+1}^2}{(t_{k+1}-u)^3}
+{\psi_k^L}'(\widehat{x}_k)+{\psi_k^R}'(\widehat{x}_k)\\
0 &> 2\frac{w_kt_k^2}{(t_k-\widehat{x}_k)^3}
+2\frac{w_{k+1}t_{k+1}^2}{(t_{k+1}-x^*_k)^3}
+{\psi_k^L}'(\widehat{x}_k)+{\psi_k^R}'(\widehat{x}_k)\\
 -2\frac{w_kt_k^2}{(t_k-\widehat{x}_k)^3}
-{\psi_k^L}'(\widehat{x}_k)-{\psi_k^R}'(\widehat{x}_k)
&>2\frac{w_{k+1}t_{k+1}^2}{(t_{k+1}-x^*_k)^3}\\
t_{k+1}-x^*_k &> \left(\frac{2w_{k+1}t_{k+1}^2}{\displaystyle-2\frac{w_kt_k^2}{(t_k-\widehat{x}_k)^3}
-{\psi_k^L}'(\widehat{x}_k)-{\psi_k^R}'(\widehat{x}_k)}\right)^{1/3}\\
x^*_k &< t_{k+1}-\left(\frac{2w_{k+1}t_{k+1}^2}{\displaystyle-2\frac{w_kt_k^2}{(t_k-\widehat{x}_k)^3}
-{\psi_k^L}'(\widehat{x}_k)-{\psi_k^R}'(\widehat{x}_k)}\right)^{1/3}\\
x^*_k &< t_{k+1}-\left(\frac{2w_{k+1}t_{k+1}^2}{\displaystyle-2\frac{w_kt_k^2}{(t_k-\widehat{x}_k)^3}
-\varphi'(\widehat{x}_k)}\right)^{1/3},
\end{align*}  
where the last inequality follows from $\theta_k'(\widehat{x}_k)=0$. Thus if we set
$$\overline{x} \defeq t_{k+1}-\left(\frac{2w_{k+1}t_{k+1}^2}{\displaystyle-2\frac{w_kt_k^2}{(t_k-\widehat{x}_k)^3}-\varphi'(\widehat{x}_k)}\right)^{1/3},$$
we know that $\varphi'(\overline{x})>0$. Launching a zero-finding algorithm for $\varphi'(\cdot)$ on the interval $[\underline{x},\overline{x}]$ as defined above yields a unique solution $x^*_k$.
\item $\varphi'(\widehat{x}_k)<0$: A similar line of reasoning points us to 
$$\underline{x}\defeq t_{k}+\left(\frac{2w_{k}t_{k}^2}{\displaystyle2\frac{w_{k+1}t_{k+1}^2}{(t_{k+1}-\widehat{x}_k)^3}+\varphi'(\widehat{x}_k)}\right)^{1/3},$$
$\overline{x}=\widehat{x}_k$, and yields a unique zero $x^*_k$ for $\varphi'(\cdot)$ over the interval $[\underline{x},\overline{x}]$.
\end{itemize}
Across all three cases, the outcome of this procedure is $x^*_k =
\mbox{argmin}_{u\in(t_{k},t_{k+1})}\varphi(u)$. Spectral separation occurs between $t_k$ and $t_{k+1}$ if and only if $\varphi(x^*_k)<1/c$.

If there is no spectral separation, then we can dismiss the interval $(t_{k},t_{k+1})$. Otherwise we need some additional work to compute spectrum boundaries.

\subsubsection{Interval Boundaries}

Consider now some $k\in\{1,2,\ldots,K-1\}$ for which
$x^*_k=\mbox{argmin}_{u\in(t_{k},t_{k+1})}\varphi(u)$ is known and
$\varphi(x^*_k)<1/c$. Spectral separation means that the support ends
at some point in $(t_{k},x^*_k)$ and starts again at some point in
$(x^*_k,t_{k+1})$. The equation that characterizes support endpoints
is  $\varphi(x)=1/c$. Thus we need to find two zeros of the function
$\varphi(\cdot)-1/c$, one in the interval $(t_{k},x^*_k)$ and the
other in the interval $(x^*_k,t_{k+1})$.

Let us start with the first zero of the function $\varphi(\cdot)-1/c$, the
one that lies in the interval $(t_{k},x^*_k)$. Once again, we employ
an off-the-shelf univariate zero-finding routine that takes as inputs
two points $(\underline{x},\overline{x})$ such that
$\varphi(\underline{x})>1/c$ and $\varphi(\overline{x})<1/c$. The
obvious candidate for $\overline{x}$ is $\overline{x} \defeq x^*_k$. For $\underline{x}$, however, we cannot use $t_k$ because $\lim_{x\searrow t_{k}}\varphi(x)=+\infty$. Therefore we need to find some $\underline{x}\in(t_{k},x^*_k)$ that verifies $\varphi(\underline{x})>1/c$. Such an $\underline{x}$ can be found by considering the following series of inequalities, which hold for all $x\in(t_k,x^*_k)$:
\begin{align*}
\varphi(x) &> \frac{w_kt_k^2}{(t_k-x)^2}
+\sum_{j=1}^{k-1}\frac{w_jt_j^2}{(t_j-x^*_k)^2}
+\sum_{j=k+1}^{K}\frac{w_jt_j^2}{(t_j-t_k)^2}\\
\varphi(x)-\varphi(x^*_k) &> 
\frac{w_kt_k^2}{(t_k-x)^2}
-\frac{w_kt_k^2}{(t_k-x^*_k)^2}
+\sum_{j=k+1}^{K}\frac{w_jt_j^2}{(t_j-t_k)^2}
-\sum_{j=k+1}^{K}\frac{w_jt_j^2}{(t_j-x^*_k)^2}\\
\varphi(x)-\frac{1}{c}&>
\frac{w_kt_k^2}{(t_k-x)^2}
-\frac{w_kt_k^2}{(t_k-x^*_k)^2}
+\left[\varphi(x^*_k)-\frac{1}{c}\right]
+\sum_{j=k+1}^{K}\frac{w_jt_j^2}{(t_j-t_k)^2}
-\sum_{j=k+1}^{K}\frac{w_jt_j^2}{(t_j-x^*_k)^2}.
\end{align*}
Notice that if we set 
$$\underline{x} \defeq t_k+\sqrt{\frac{w_kt_k^2}{\displaystyle\frac{w_kt_k^2}{(t_k-x^*_k)^2}
+\left[\frac{1}{c}-\varphi(x^*_k)\right]
+\sum_{j=k+1}^{K}\frac{w_jt_j^2}{(t_j-x^*_k)^2}
-\sum_{j=k+1}^{K}\frac{w_jt_j^2}{(t_j-t_k)^2}}}~,$$
then 
$$\frac{w_kt_k^2}{(t_k-\underline{x})^2}
-\frac{w_kt_k^2}{(t_k-x^*_k)^2}
+\left[\varphi(x^*_k)-\frac{1}{c}\right]
+\sum_{j=k+1}^{K}\frac{w_jt_j^2}{(t_j-t_k)^2}
-\sum_{j=k+1}^{K}\frac{w_jt_j^2}{(t_j-x^*_k)^2}=0 ~;$$
therefore, $\varphi(\underline{x})>1/c$. Feeding $(\underline{x},\overline{x})$ thus defined into the zero-finding numerical routine with the function $\varphi(\cdot)-1/c$ yields an endpoint of the support.

A similar line of reasoning leads to setting $\underline{x} \defeq x^*_k$, 
$$\overline{x} \defeq t_{k+1}-\sqrt{\frac{w_{k+1}t_{k+1}^2}{\displaystyle\frac{w_{k+1}t_{k+1}^2}{(t_{k+1}-x^*_k)^2}
+\left[\frac{1}{c}-\varphi(x^*_k)\right]
+\sum_{j=1}^{k-1}\frac{w_jt_j^2}{(t_j-x^*_k)^2}
-\sum_{j=1}^{k-1}\frac{w_jt_j^2}{(t_j-t_{k+1})^2}}}~,$$
and running a numerical routine to find a zero of the function $\varphi(\cdot)-1/c$ on the interval $(\underline{x},\overline{x})\subset(x^*_k,t_{k+1})$. This zero will also be a support endpoint.

\subsection{Extremities of the Support}
	
The procedure described so far identifies all support endpoints lying in the interval $[t_1,t_K]$. In order to complete the determination of the support, we must find the support endpoint that lies in the interval $(-\infty,t_1)$ and the support endpoint that lies in the interval $(t_K,+\infty)$.

\subsubsection{Minimum of the Support}

Let us start with the first support endpoint, the one lying in the interval $(-\infty,t_1)$. The equation that characterizes this point is the same as before: $\varphi(x)=1/c$. In order to employ the zero-finding numerical routine, we must find two bounds $\underline{x}$ and $\overline{x}$, both strictly less than~$t_1$, such~that $\varphi(\underline{x})<1/c$ and $\varphi(\overline{x})>1/c$. The left-hand side bound $\underline{x}$ can be obtained by considering the following inequalities:
\begin{align}
\forall x\in(-\infty,t_1)\quad\forall j=1,\dots,K\qquad 
\frac{w_jt_j^2}{(x-t_j)^2}&\leq\frac{w_jt_j^2}{(x-t_1)^2}\nonumber\\
\forall x\in(-\infty,t_1)\qquad
\varphi(x)&\leq\frac{\sum_{j=1}^Kw_jt_j^2}{(x-t_1)^2}~.\label{eq:lobound}
\end{align}
Notice that if we set 
$$\underline{x} \defeq t_1-\sqrt{c\sum_{j=1}^Kw_jt_j^2}-1~,$$
then 
$$\frac{\sum_{j=1}^Kw_jt_j^2}{(\underline{x}-t_1)^2}<\frac{1}{c}~,$$
which in turn implies by equation (\ref{eq:lobound}) that $\varphi(\underline{x})<1/c$, as desired.

The right-hand side bound $\overline{x}$ can be found by considering a different inequality:
\be
\forall x\in(-\infty,t_1)\qquad
\varphi(x)\geq\frac{w_1t_1^2}{(x-t_1)^2}~.\label{eq:hibound}
\ee
Notice that if we set 
$$\overline{x} \defeq t_1-\frac{\sqrt{cw_1t_1^2}}{2}~,$$
then 
$$\frac{w_1t_1^2}{(\overline{x}-t_1)^2}>\frac{1}{c}~,$$
which in turn implies by equation (\ref{eq:hibound}) that $\varphi(\overline{x})>1/c$, as desired. Launching the numerical routine to find a zero of the function $\varphi(\cdot)-1/c$ over the interval $(\underline{x},\overline{x})$ thus defined yields the first endpoint of the support.

\subsubsection{Maximum of the Support}

For the last endpoint of the support, the one that lies in the interval $(t_K,+\infty)$, a similar line of reasoning leads us to define:
\begin{align*}
\underline{x} & \defeq  t_K+\frac{\sqrt{cw_Kt_K^2}}{2}\\
\mbox{and}\qquad\overline{x}  & \defeq  t_K+\sqrt{c\sum_{j=1}^Kw_jt_j^2}+1~.
\end{align*}
Launching the numerical routine to find a zero of the function $\varphi(\cdot)-1/c$ over the interval $(\underline{x},\overline{x})$ thus defined yields the last endpoint of the support.

\subsection{Output}

The main outputs of this procedure are $\nu\geq1$, the number of distinct intervals that constitute the support, and $u_1,\ldots,u_{2\nu}$, the support endpoints. The support in $u$-space is $S_U=[u_1,u_2]\cup\dots\cup[u_{2\nu-1},u_{2\nu}]$. 

Another output of this procedure is a set of positive integers $\omega_1,\ldots,\omega_\nu$ summing up to $p$ that tell us how many population eigenvalues correspond to each support interval. If $\nu=1$ then there is no spectral separation and $\omega_1=p$. If $\nu\geq2$ and the first spectral separation occurs between $t_k$ and $t_{k+1}$ for some $k=1,\ldots,K-1$, then $\omega_1=p\sum_{j=1}^kw_j$. If some poulation eigenvalues are equal to zero then $\omega_1$ needs to be augmented accordingly.

If $\nu\geq2$ and the last  spectral separation occurs between $t_{k'}$ and $t_{k'+1}$ for some $k'=1,\ldots,K-1$, then $\omega_\nu=p\sum_{j=k'+1}^Kw_j$. If $\nu\geq3$ and the $i^{\rm th}$ support interval (for $i=2,\ldots,\nu-1$) is delimited on the left-hand side by spectral separation occurring between $t_k$ and $t_{k+1}$, and on the right-hand side by spectral separation occurring between $t_{k'}$ and $t_{k'+1}$ (where $1\leq k < k' \leq K$), then $\omega_i=p\sum_{j=k+1}^{k'}w_j$. This information will turn out to be useful in subsequent operations.

\subsection{Derivative of the Support Endpoints}
\label{sec:supgrd}

If the QuEST function defined by equations (\ref{eq:population})--(\ref{eq:questMP}) is to be used efficiently in an optimization algorithm, it is desirable to be able to compute its derivative analytically. Since this function is constructed as a chain of six successive operations, the first of which is the determination of support endpoints, its derivative can be computed in the same way, provided that we start by computing analytically the derivative of support endpoints with respect to $\tau_k$ for all $k=1,\ldots,K$.

Every $u_i$ for $i=1,\ldots,2\nu$ is a zero of the function
$$\widetilde{\varphi}(u;\tau_1,\ldots,\tau_p) \defeq \frac{1}{p}\sum_{j=1}^p\frac{\tau_j^2}{(\tau_j-u)^2}-\frac{1}{c}~.$$
By differentiating the equation $\widetilde{\varphi}(u;\tau_1,\ldots,\tau_p)=0$ we get:
\begin{align*}
\frac{\partial\widetilde{\varphi}}{\partial u}\cdot du
+\frac{\partial\widetilde{\varphi}}{\partial \tau_k}\cdot d\tau_k&=0~,\\
\frac{\partial u}{\partial \tau_k}&=-\frac{\displaystyle\frac{\partial\widetilde{\varphi}}{\partial \tau_k}}{\displaystyle\frac{\partial\widetilde{\varphi}}{\partial u}}~.
\end{align*}
The partial derivatives of the function $\widetilde{\varphi}$ are as follows:
\begin{align*}
\frac{\partial\widetilde{\varphi}}{\partial u}(u;\tau_1,\ldots,\tau_p)&=\frac{2}{p}\sum_{j=1}^p\frac{\tau_j^2}{(\tau_j-u)^3}\\
\frac{\partial\widetilde{\varphi}}{\partial \tau_k}(u;\tau_1,\ldots,\tau_p)
&=-\frac{2}{p}\frac{\tau_k u}{(\tau_k-u)^3}~;
\end{align*}
therefore,
\be
\forall i=1,\ldots,2\nu\quad \forall k=1,\ldots,p\qquad
\frac{\partial u_i}{\partial \tau_k}=
\frac{\displaystyle\frac{\tau_k u_i}{(\tau_k-u_i)^3} }{\displaystyle\sum_{j=1}^p\frac{\tau_j^2}{(\tau_j-u_i)^3}}~.\label{eq:supgrd}
\ee

\section{Grid}

The first operation generated the support in $u$-space $S_U=[u_1,u_2]\cup\dots\cup[u_{2\nu-1},u_{2\nu}]$ and the number of population eigenvalues corresponding to each interval: $\omega_1,\ldots,\omega_\nu$. The goal of the second operation is to produce a grid that covers this support. This problem can be broken down by considering each interval $i=1,\ldots,\nu$ separately.

\subsection{Formula for the Grid Points}
\label{sub:grid}

Take some $i\in\{1,\ldots,\nu\}$. How shall we determine a grid that
covers the interval $[u_{2i-1},u_{2i}]$? The number of points on the
grid will be a function of $\omega_i$. Specifically, we shall take
$\omega_i$~points in the open interval $(u_{2i-1},u_{2i})$, plus the
two endpoints $u_{2i-1}$ and $u_{2i}$. Thus, the total number of
points covering the closed interval $[u_{2i-1},u_{2i}]$ will be
$\omega_i+2$. Let us call these points
$\xi_0^i,\ldots,\xi_{\omega_i+1}^i$, with the convention that $\xi_0^i
\defeq u_{2i-1}$ and $\xi_{\omega_i+1}^i \defeq u_{2i}$. Thus, what is left is to define $\xi_1^i,\ldots,\xi_{\omega_i}^i$.

There are many ways to choose such a grid, depending on how densely we want to cover the various parts of the interval. The simplest idea would be to have uniform coverage through a linearly spaced grid. But it is more judicious to increase coverage density near the edges of the interval because this is where a lot of the action is taking place. \cite{silverstein:choi:1995} demonstrate that the limiting density of sample eigenvalues has ``square root''-type behavior near boundary points. This fact points us towards the inverse c.d.f.~function of the beta distribution with parameters $(0.5,0.5)$, also known as the {\em arcsine} distribution:
\be
\forall j\in\{0,\ldots,\omega_i+1\}\qquad \xi_j^i \defeq u_{2i-1}+(u_{2i}-u_{2i-1})\,
\sin^2\left[\frac{\pi j}{2(\omega_i+1)}\right]~.\label{eq:netfun}
\ee
Compared to the beta distribution with parameters $(1,1)$, which is the uniform distribution, reducing both parameters from $1$ to $0.5$ increases coverage density near the edges of the interval. Note that the density of the arcsine distribution goes to infinity at the edges of the interval (as does the derivative of the square root function), but the c.d.f.,~its inverse and the grid all remain well-behaved. The goal here is to enhance numerical accuracy.

\subsection{Derivative of the Grid Points}
\label{sub:netgrd}

In keeping with our earlier stated objective (see Section \ref{sec:supgrd}) of building towards an analytical formula for the partial derivative of $\lambda_i$ with respect to $\tau_k$ for all $i,k\in\{1,\ldots,p\}$, at this stage we need to compute $\partial\xi_j^i/\partial\tau_k$ for all $j\in\{1,\ldots,\omega_i\}$. From equation (\ref{eq:netfun}) we can see immediately that it is
\be
\frac{\partial\xi_j^i}{\partial\tau_k}=\left\{1-\sin^2\left[\frac{\pi j}{2(\omega_i+1)}\right]\right\}\frac{\partial u_{2i-1}}{\partial\tau_k}
+\sin^2\left[\frac{\pi j}{2(\omega_i+1)}\right]\frac{\partial u_{2i}}{\partial\tau_k}~,\label{eq:netgrd}
\ee
where $\partial u_{2i-1}/\partial\tau_k$ and $\partial u_{2i}/\partial\tau_k$ are given by equation (\ref{eq:supgrd}).

\section{Solving the Mar\v{c}enko-Pastur Equation in {\boldmath $u$}-Space}

In this section we will assume that the interval index $i\in\{1,\ldots,\nu\}$ is fixed.

\subsection{Statement of the Problem}
\label{sub:solfun}

Given a grid coverage $(\xi_j^i)_{j=0,\ldots,\omega_i}$ of the $i^{\rm th}$ support interval, the third operation solves the Mar\v{c}enko-Pastur equation at $\xi_j^i$. For every $j=0,\ldots,\omega_i+1$, define the function
$$\forall y\in[0,+\infty) \qquad \Gamma_j^i(y) \defeq \frac{1}{p}\sum_{k=1}^p\frac{\tau_k^2}{(\tau_j-\xi_j^i)^2+y^2}-\frac{1}{c}.$$
It is easy to verify that $\Gamma_j^i$ is strictly decreasing on $[0,+\infty)$ and that $\lim_{y\to+\infty}\Gamma_j^i(y)=-1/c$. 

The solution to the Mar\v{c}enko-Pastur equation at $\xi_j^i$ is the unique $y\in[0,+\infty)$ such that
\be
\Gamma_j^i(y)=0~.\label{eq:gamma}
\ee
Call it $y_j^i$. This line of attack is directly inspired by Section 2.3 of \cite{ledoit:wolf:2012}. 

From the definition of the $\xi_j^i$'s in Section \ref{sub:grid}, it is obvious that $y_1^i=y_{\omega_i+1}^i=0$. What remains to be determined is $(y_j^i)_{j=1,\ldots,\omega_i}$. In the remainder of this section we will assume that $j$ is fixed in the set $\{1,\ldots,\omega_i\}$.

The solution $y$ to the equation $\Gamma_j^i(y)=0$ is computed by some standard numerical routine that finds the zero of a real univariate function. As usual, we need to input into this routine a lower bound $\underline{y}_j^i\in[0,+\infty)$ such that $\Gamma_j^i(\underline{y}_j^i)\in(0,+\infty)$ and an upper bound $\overline{y}_j^i\in(0,+\infty)$ such that $\Gamma_j^i(\overline{y}_j^i)<0$.

\subsection{Lower Bound}

From Section \ref{sec:support}, $(t_1,\ldots,t_K)$ is the vector of
unique nonzero population eigenvalues, with corresponding weights
$(w_1,\ldots,w_K)$. Let $\delta_j^i \defeq
\min_{k\in\{1,\ldots,K\}}(t_k-\xi_i^j)^2$ and $\Omega_i^j \defeq 
\bigl \{k\in\{1,\ldots,K\}:(t_k-\xi_i^j)^2=\delta_j^i \bigr \}$. Then we have
\be
\Gamma_j^i(y)\geq\frac{\sum_{k\in\Omega_i^j}w_k t_k^2}{\delta_j^i+y^2}-\frac{1}{c}~.\label{eq:lower}
\ee
Looking at the right-hand side of equation (\ref{eq:lower}), we see that
$$\frac{\sum_{k\in\Omega_i^j}w_k t_k^2}{\delta_j^i+y^2}-\frac{1}{c}\geq 0 \Longleftrightarrow
y^2\leq c\sum_{k\in\Omega_i^j}w_k t_k^2-\delta_j^i~.$$
Therefore, if we set
$$\underline{y}_j^i \defeq \frac{\sqrt{\max\left(0,c\sum_{k\in\Omega_i^j}w_k t_k^2-\delta_j^i\right)}}{2}~,$$
then $\Gamma(\underline{y}_j^i)\in(0,+\infty)$, as desired.

\subsection{Upper Bound}

We use the inequalities
\begin{align}
\forall k\in\{1,\ldots,K\}
\qquad\frac{1}{(t_k-\xi_j^i)^2+y^2}&\leq \frac{1}{\delta_j^i+y^2}\nonumber\\
\Gamma_j^i(y)&\leq\frac{\sum_{k=1}^Kw_kt_k^2}{\delta_j^i+y^2}-\frac{1}{c}~.\label{eq:higher}
\end{align}
Notice that if we set
$$\overline{y}_j^i \defeq \sqrt{c\sum_{k=1}^Kw_kt_k^2-\delta_j^i}+1~,$$
then 
$$\frac{\sum_{k=1}^Kw_kt_k^2}{\delta_j^i+(\overline{y}_j^i)^2}-\frac{1}{c}<0~;$$
therefore, by equation (\ref{eq:higher}), $\Gamma_j^i(\overline{y}_j^i)<0$, as desired.

\subsection{Output}

Launching a standard numerical routine to find the zero of the
function $\Gamma_j^i(\cdot)$ over the interval
$(\underline{y}_j^i,\overline{y}_j^i)$ yields $y_j^i$, the solution to
the Mar\v{c}enko-Pastur equation at $\xi_j^i$. The output of this
operation is more conveniently expressed as the complex number $z_j^i
\defeq \xi_j^i+\sqrt{-1}\,y_j^i$.

\subsection{Derivative}

The derivative of the real part of $z_j^i$ with respect to $\tau_k$ has been computed in Section \ref{sub:netgrd}. As~for the derivative of the imaginary part, $y_j^i$, consider the function
$$\widetilde{\Gamma}_j^i(y;\tau_1,\ldots,\tau_p) \defeq \frac{1}{p}\sum_{k=1}^p\frac{\tau_k^2}{\left(\tau_k-\xi_j^i\right)^2+y^2}-\frac{1}{c}~.$$
We can view $y_j^i$ as a function of $(\tau_1,\ldots,\tau_p)$: $y_j^i=\widetilde{y}_j^i(\tau_1,\ldots,\tau_p)$. Then the manner in which $y_j^i$ is obtained in Section \ref{sub:solfun} can be expressed through the equation
$$\widetilde{\Gamma}_j^i\left(\widetilde{y}_j^i(\tau_1,\ldots,\tau_p);\tau_1,\ldots,\tau_p\right)=0.$$
Taking the partial derivative with respect to $\tau_k$ while holding the other population eigenvalues constant yields
\begin{align*}
\frac{\partial\widetilde{\Gamma}_j^i}{\partial y}
\cdot\frac{\partial\widetilde{y}_j^i}{\partial \tau_k}
+\frac{\partial\widetilde{\Gamma}_j^i}{\partial \tau_k}&=0~,\\
\frac{\partial\widetilde{y}_j^i}{\partial \tau_k}&=
-\frac{\displaystyle\frac{\partial\widetilde{\Gamma}_j^i}{\partial \tau_k}}{\displaystyle
\frac{\partial\widetilde{\Gamma}_j^i}{\partial y}}~.
\end{align*}
The partial derivatives of the function $\widetilde{\Gamma}_j^i$ are
\begin{align*}
\frac{\partial\widetilde{\Gamma}_j^i}{\partial \tau_k}(y;\tau_1,\ldots,\tau_p) &=
\frac{2\tau_k}{\left(\tau_k-\xi_j^i\right)^2+y^2}
-\frac{2\tau_k^2\left(\tau_k-\xi_j^i\right)}{\left[\left(\tau_k-\xi_j^i\right)^2+y^2\right]^2}\\
\frac{\partial\widetilde{\Gamma}_j^i}{\partial y}(y;\tau_1,\ldots,\tau_p) &=
-2\sum_{l=1}^p\frac{\tau_l^2y}{\left[\left(\tau_l-\xi_j^i\right)^2+y^2\right]^2}~.
\end{align*}
Therefore,
\be
\frac{\partial\widetilde{y}_j^i}{\partial \tau_k}(\tau_1,\ldots,\tau_p)=
\frac{\displaystyle
\frac{\tau_k}{(\tau_k-\xi_j^i)^2+(y_j^i)^2}
-\frac{\tau_k^2(\tau_k-\xi_j^i)}{[(\tau_k-\xi_j^i)^2+(y_j^i)^2]^2}}{\displaystyle
\sum_{l=1}^p\frac{\tau_l^2\,y_j^i}{[(\tau_l-\xi_j^i)^2+(y_j^i)^2]^2}}~.\label{eq:dydtau}
\ee
Now this is only part of the answer because in this analysis we held $\xi_j^i$ constant, whereas in reality it is also a function of the population eigenvalues. Thus, the partial derivative of $y_j^i$ with respect to $\tau_k$ is given by the formula
\be
\frac{\partial y_j^i}{\partial\tau_k}=
\frac{\partial\widetilde{y}_j^i}{\partial \tau_k}
+\frac{\partial y_j^i}{\partial\xi_j^i}
\cdot 
\frac{\partial\xi_j^i}{\partial\tau_k}~,\label{eq:solgrd}
\ee
where $\partial\widetilde{y}_j^i/\partial \tau_k$ is given by equation (\ref{eq:dydtau}) and $\partial\xi_j^i/\partial\tau_k$ is given by equation (\ref{eq:netgrd}). All that remains to be computed is $\partial y_j^i/\partial\xi_j^i$. This is done by temporarily ignoring direct dependency on population eigenvalues and setting up the function
$$\widehat{\Gamma}(y;\xi) \defeq \frac{1}{p}\sum_{k=1}^p\frac{\tau_k^2}{\left(\tau_j-\xi\right)^2+y^2}-\frac{1}{c}~.$$
Differentiating the equation $\widehat{\Gamma}(y;\xi)=0$ yields:
$$\frac{\partial\widehat{\Gamma}}{\partial y}\,dy+
\frac{\partial\widehat{\Gamma}}{\partial \xi}\,d\xi=0\quad
\Longrightarrow\quad
\frac{\partial y}{\partial\xi}=-\frac{\displaystyle\frac{\partial\widehat{\Gamma}}{\partial \xi}}{\displaystyle\frac{\partial\widehat{\Gamma}}{\partial y}}~.$$
The partial derivatives of the function $\widehat{\Gamma}$ are
\begin{align*}
\frac{\partial\widehat{\Gamma}}{\partial \xi}(y;\xi) &= 2\sum_{l=1}^p\frac{\tau_l^2(\tau_l-\xi)}{\left[\left(\tau_l-\xi\right)^2+y^2\right]^2}\\
\mbox{and}\qquad
\frac{\partial\widehat{\Gamma}}{\partial y}(y;\xi) &= -2\sum_{l=1}^p\frac{\tau_l^2\,y}{\left[\left(\tau_l-\xi\right)^2+y^2\right]^2}~;
\end{align*}
therefore,
$$\frac{\partial y_j^i}{\partial\xi_j^i}=
\frac{\displaystyle\sum_{l=1}^p\frac{\tau_l^2(\tau_l-\xi_j^i)}{\left[\left(\tau_l-\xi_j^i\right)^2
+\left(y_j^i\right)^2\right]^2}}{\displaystyle
\sum_{l=1}^p\frac{\tau_l^2\,y_j^i}{\left[\left(\tau_l-\xi_j^i\right)^2
+\left(y_j^i\right)^2\right]^2}}~.$$
Plugging this formula into equation (\ref{eq:solgrd}) yields the partial derivative of $y_j^i$ with respect to $\tau_k$.

\section{\mbox{Density of the Limiting Distribution of the Sample Eigenvalues}}

\subsection{Mapping}

This is the operation where we leave $u$-space and map back to $(x,F(x))$ where $F$ is the limiting distribution of sample eigenvalues. The underlying mathematics for this mapping can be found in equations (2.7)--(2.8) of \cite{ledoit:wolf:2012}. The mapping can be expressed with the notation of the present paper as
$$x \defeq u-c\,u\,\frac{1}{p}\sum_{k=1}^p\frac{\tau_k}{\tau_k-u}~.$$
In the remainder of this section, we will assume that the interval index $i\in\{1,\ldots,\nu\}$ is fixed. For every $j\in\{0,1,\ldots,\omega_i+1\}$, map $z_j^i$ into:
\be
x_j^i=z_j^i-c\,z_j^i\,\frac{1}{p}\sum_{k=1}^p\frac{\tau_k}{\tau_k-z_j^i}~.\label{eq:denfun:x}
\ee
Even though $z_j^i$ is generally a complex number, equation (\ref{eq:gamma}) guarantees that $x_j^i$ is real.

Using Section (2.3) of \cite{ledoit:wolf:2012}, we can also obtain the value of the limiting sample spectral density $F'$ evaluated at $x_j^i$ as $F'(x_j^i)=f_j^i$ where
\be
f_j^i=\frac{1}{c\pi}\im\left[-\frac{1}{z_j^i}\right]=
\frac{1}{c\pi}\frac{y_j^i}{(x_j^i)^2+(y_j^i)^2}~.\label{eq:denfun:f}
\ee
Note that $f_1^i=f_{\omega_i+1}^i=0$.

The output of this operation is $(x_j^i,f_j^i)_{j=0,1,\ldots,\omega_i+1}$, for every $i\in\{1,\ldots,\nu\}$.

\subsection{Derivative}

From equation (\ref{eq:denfun:f}), it is easy to compute the partial derivative of $f_j^i$ with respect to $\tau_k$ as
\begin{align}
\frac{\partial f_j^i}{\partial\tau_k} &=\frac{1}{c\pi}\,\im\left[
\frac{\partial z_j^i}{\partial\tau_k}\cdot\frac{1}{\left(z_j^i\right)^2}\right]~,\label{eq:dengrd:f}\\
\mbox{where}\qquad\frac{\partial z_j^i}{\partial\tau_k} &=\frac{\partial x_j^i}{\partial\tau_k}+\sqrt{-1}\,\frac{\partial y_j^i}{\partial\tau_k}~,\label{eq:dz}
\end{align}
$\partial x_j^i/\partial\tau_k$ is given by equation (\ref{eq:netgrd}), and 
$\partial y_j^i/\partial\tau_k$ is given by equation (\ref{eq:solgrd}).

In order to differentiate equation (\ref{eq:denfun:x}) more easily, introduce the function $m_{LH}$ defined as per Section 2.2 of \cite{ledoit:wolf:2012}:
$$\forall z\in\C^+\qquad m_{LH}(z;\tau_1,\ldots,\tau_p) \defeq \frac{1}{p}\sum_{l=1}^p\frac{\tau_l}{\tau_l-z}
=1+z\,\frac{1}{p}\sum_{l=1}^p\frac{1}{\tau_l-z}~.$$
This enables us to rewrite equation (\ref{eq:denfun:x}) as 
\be
x_j^i=z_j^i-c\,z_j^i\,m_{LH}(z_j^i;\tau_1,\ldots,\tau_p)~.\label{eq:mLH}
\ee
The full derivative of $m_{LH}(z_j^i;\tau_1,\ldots,\tau_p)$ with respect to $\tau_k$ is
$$\frac{dm_{LH}}{d\tau_k}=\frac{\partial m_{LH}}{\partial\tau_k}
+\frac{\partial m_{LH}}{\partial z_j^i}\cdot\frac{\partial z_j^i}{\partial\tau_k}~,$$
where the last term is given by equation (\ref{eq:dz}). The partial derivatives of $m_{LH}$ are
\begin{align*}
\frac{\partial m_{LH}}{\partial\tau_k} &= -z_j^i\times\frac{1}{p}\frac{1}{(\tau_k-z_j^i)^2}\\
\mbox{and}\qquad\frac{\partial m_{LH}}{\partial z_j^i} &= \frac{1}{p}\sum_{l=1}^p\frac{\tau_l}{(\tau_l-z_j^i)^2}~;
\end{align*}
therefore,
$$\frac{dm_{LH}}{d\tau_k}
=-z_j^i\times\frac{1}{p}\frac{1}{(\tau_k-z_j^i)^2}
+\frac{\partial z_j^i}{\partial\tau_k}
\times \frac{1}{p}\sum_{l=1}^p\frac{\tau_l}{(\tau_l-z_j^i)^2}~.$$
Finally, differentiating equation (\ref{eq:mLH}) enables us to compute the partial derivative of $x_j^i$ with respect to $\tau_k$ as follows:
\be
\frac{\partial x_j^i}{\partial\tau_k} 
=\frac{\partial z_j^i}{\partial\tau_k} 
\times\left[1-c\cdot m_{LH}(z_j^i;\tau_1,\ldots,\tau_p)\right]
-c\cdot z_j^i\,\frac{dm_{LH}}{d\tau_k}(z_j^i;\tau_1,\ldots,\tau_p)~.\label{eq:dengrd:x}
\ee

\section{Cumulative Distribution Function}

\subsection{Numerical Integration of the Density}
\label{sub:cdf}

The objective is to compute $F_j^i \defeq F(x_j^i)$.
We know that 
\be
F(0)=F_0^1=\max\left(0,1-\frac{1}{c}\right)~.\label{eq:null}
\ee
Since the support of $F$ is 
$\cup_{i=1}^\nu[x^i_{0},x^i_{\omega_i+1}]$ (with  the possible addition of $\{0\}$ if $p>n$), as soon as $\nu$ is greater than or equal to two,
$F^{i+1}_{0}=F^i_{\omega_i+1}$, for $i=1,\ldots,\nu-1$.
\cite{bai:silverstein:1999} show that
\be
\forall i=1,\ldots,\nu\qquad
F^i_{\omega_i+1}=\frac{1}{p}\sum_{j=1}^i\omega_j~.\label{eq:interval}
\ee
 All that remains is to compute 
$F^i_j$ for $j\in\{1,\ldots,\omega_i\}$. First, we will get an approximation of~$F^i_j$ by using the trapezoidal integration formula over $[x^i_0,x^i_j]$. Then we will refine this approximation using the fact stated in equation (\ref{eq:interval}). The trapezoidal method yields the approximation:
\be
\forall j=1,\ldots,\omega_i+1
\qquad \widetilde{F}^i_j \defeq F^i_0
+\frac{1}{2}
\sum_{l=1}^j(x^i_{l}-x^i_{l-1})(f^i_{l}+f^i_{l-1})~.\label{eq:trapeze}
\ee
Now the problem is that $\widetilde{F}^i_{\omega_i+1}$ thus defined would generally differ from $\sum_{j=1}^i\omega_j/p$ due to numerical error in the integration formula. This is why, in a second step, we refine the approximation by computing
\be
F^i_{j} \defeq F^i_0+\left(\widetilde{F}^i_j-F^i_0\right)
\frac{F^i_{\omega_i+1}-F^i_0}{\widetilde{F}^i_{\omega_i+1}-F^i_0}
\qquad\mbox{for}\quad j=1,\ldots,\omega_i~.\label{eq:disfun}
\ee

\subsection{Derivatives with Respect to Population Eigenvalues}

The computation of these derivatives is subdivided into two steps that mirror the ones performed in Section \ref{sub:cdf}. First, by differentiating equation (\ref{eq:trapeze}) with respect to $x^i_{l}$ and $f^i_{l}$ we~obtain
\begin{align}
\forall j=1,\ldots,\omega_i+1
\qquad \frac{\partial\widetilde{F}^i_j}{\partial \tau_k}&=\frac{1}{2}
\sum_{l=1}^j\left(\frac{\partial x^i_{l}}{\partial\tau_k}
-\frac{\partial x^i_{l-1}}{\partial\tau_k}\right)\left(f^i_{l}+f^i_{l-1}\right)\nonumber\\
&\quad+\frac{1}{2}\sum_{l=1}^j\left(x^i_{l}-x^i_{l-1}\right)
\left(\frac{\partial f^i_{l}}{\partial\tau_k}+\frac{\partial f^i_{l-1}}{\partial\tau_k}\right),\label{eq:dftilde}
\end{align}
where the partial derivatives of $x^i_{l}$ and $f^i_{l}$ with respect to $\tau_k$ are given by equations (\ref{eq:dengrd:x}) and~(\ref{eq:dengrd:f}), respectively. Second, differentiating equation (\ref{eq:disfun}) with respect to $\widetilde{F}^i_j$ and $\widetilde{F}^i_{\omega_i+1}$ yields
\be
\frac{\partial F^i_{j}}{\partial\tau_k}= \left(F^i_{\omega_i+1}-F^i_0\right)\frac{\displaystyle\frac{\partial\widetilde{F}^i_j}{\partial\tau_k}-F^i_0}{\widetilde{F}^i_{\omega_i+1}-F^i_0}
-\left(F^i_{\omega_i+1}-F^i_0\right)\frac{\partial\widetilde{F}^i_{\omega_i+1}}{\partial\tau_k}
\cdot\frac{\widetilde{F}^i_j-F^i_0}{\left(\widetilde{F}^i_{\omega_i+1}-F^i_0\right)^2}~,
\label{eq:disgrd}
\ee
where the partial derivatives of $\widetilde{F}^i_j$ and $\widetilde{F}^i_{\omega_i+1}$ with respect to $\tau_k$ are given by equation (\ref{eq:dftilde}).

\section{Discretization of the Sample Spectral C.D.F.}
\label{sec:final}

\subsection{Sample Eigenvalues}
\label{sub:intfun}

The final operation involves extracting from $F$ a set of $p$ sample eigenvalues $(\lambda_1,\ldots,\lambda_p)$. First, we take care of zero eigenvalues when $c>1$. By equation (\ref{eq:null}) we know that
$$\mbox{if}\quad p<n\quad\mbox{then}\quad\lambda_1,\ldots,\lambda_{p-n}=0~.$$
In what follows we will assume that we have fixed an interval index $i$ in the set $\{1,\ldots,\nu\}$. 

Let the function $X^i(\alpha)$ denote the approximation to $\int_{F_0^i}^\alpha F^{-1}(x)dx$ that is obtained by fitting a piecewise linear function to $F^{-1}(\cdot)$ over the interval $[F_0^i,F_{\omega_i+1}^i]$. This piecewise linear function passes through every point $(F_j^i,x_j^i)_{j=0,\ldots,\omega_i+1}$. Using once again the trapezoidal integration formula, we get:
\begin{align}
\forall j=0,\ldots,\omega_i\qquad
\int_{F_j^i}^{F_{j+1}^i}F^{-1}(x)\,dx\approx
X^i(F_{j+1}^i)-X^i(F_j^i)=
\left(F_{j+1}^i-F_j^i\right)\frac{x_j^i+x_{j+1}^i}{2}~.\label{eq:intxdF}
\end{align}
For every integer $\kappa$ such that $pF_0^i\leq \kappa<pF_{\omega_i+1}^i$, define $j(\kappa)$ as the unique integer in $\{0,\ldots,\omega_i\}$ such that $F_{j(\kappa)}^i\leq \kappa<F_{j(\kappa)+1}^i$. Then we have:
\begin{align}
\int_{F_{j(\kappa)}^i}^{\kappa/p}F^{-1}(x)\,dx&\approx
X^i(\kappa/p)-X^i\left(F_{j(\kappa)}^i\right),\quad\mbox{where}\nonumber\\
X^i(\kappa/p)-X^i\left(F_{j(\kappa)}^i\right)
&=\left(\frac{\kappa}{p}-F_{j(\kappa)}^i\right)\left[x_{j(\kappa)}^i+
\frac{\frac{\kappa}{p}-F_{j(\kappa)}^i}{2\left(F_{j(\kappa)+1}^i-F_{j(\kappa)}^i\right)}
\left(x_{j(\kappa)+1}^i-x_{j(\kappa)}^i\right)\right]\nonumber\\
&=\left(\frac{\kappa}{p}-F_{j(\kappa)}^i\right)x_{j(\kappa)}^i
+\frac{\left(\frac{\kappa}{p}-F_{j(\kappa)}^i\right)^2}{2\left(F_{j(\kappa)+1}^i-F_{j(\kappa)}^i\right)}
\,\left(x_{j(\kappa)+1}^i-x_{j(\kappa)}^i\right)~.\label{eq:intxdF2}
\end{align}
Putting together equations (\ref{eq:intxdF})--(\ref{eq:intxdF2}) yields
\begin{align*}
X^i(\kappa/p)&=\sum_{l=0}^{j(\kappa)-1}\left(F_{l+1}^i-F_l^i\right)\frac{x_l^i+x_{l+1}^i}{2}\\
&\quad+\left(\frac{\kappa}{p}-F_{j(\kappa)}^i\right)x_{j(\kappa)}^i
+\frac{\left(\frac{\kappa}{p}-F_{j(\kappa)}^i\right)^2}{2\left(F_{j(\kappa)+1}^i-F_{j(\kappa)}^i\right)}
\,\left(x_{j(\kappa)+1}^i-x_{j(\kappa)}^i\right)~.
\end{align*}
Finally, we can define the sample eigenvalues that belong to the $i^{\rm th}$ support interval as:
\be
\forall \kappa\in\{pF_0^i+1,pF_0^i+2,\ldots,pF_{\omega_i+1}^i\}\qquad
\lambda_\kappa \defeq X^i\left(\frac{\kappa}{p}\right)-X^i\left(\frac{\kappa-1}{p}\right)~.\label{eq:intfun}
\ee

\subsection{Partial Derivatives of Sample Eigenvalues w.r.t.~Population Eigenvalues}

As in Section \ref{sub:intfun}, we handle separately the zero eigenvalues when the sample covariance matrix is singular:
$$\mbox{if}\quad p<n\quad\mbox{then}\quad\forall \kappa=1,\ldots,p-n\qquad\frac{\partial\lambda_\kappa}{\partial\tau_k}=0.$$
In the remainder of this section we will assume that we have fixed an interval index $i$ in the set $\{1,\ldots,\nu\}$. 
Differentiating equation (\ref{eq:intxdF}) with respect to $F_j^i$ and $x_j^i$ yields
\begin{align}
\forall j=0,\ldots,\omega_i\qquad
\frac{\partial X^i}{\partial\tau_k}(F_{j+1}^i)-\frac{\partial X^i}{\partial\tau_k}(F_j^i)
&=\frac{1}{2}\left(\frac{\partial F_{j+1}^i}{\partial\tau_k}-\frac{\partial F_j^i}{\partial\tau_k}\right)
\left(x_j^i+x_{j+1}^i\right)\nonumber\\
&\quad+\frac{1}{2}
\left(F_{j+1}^i-F_j^i\right)
\left(\frac{\partial x_j^i}{\partial\tau_k}+\frac{\partial x_{j+1}^i}{\partial\tau_k}\right)
~,\label{eq:dintxdF}
\end{align}
where the partial derivatives of $F_j^i$ and $x_j^i$ with respect to $\tau_k$ are given by equations~(\ref{eq:disgrd}) and~(\ref{eq:dengrd:x}), respectively. Similarly, differentiating equation (\ref{eq:intxdF2}) yields
\begin{align}
\frac{\partial X^i}{\partial\tau_k}\left(\frac{\kappa}{p}\right)-\frac{\partial X^i}{\partial\tau_k}\left(F_{j(\kappa)}^i\right)
&=\left(\frac{\kappa}{p}-F_{j(\kappa)}^i\right)
\frac{\partial x_{j(\kappa)}^i}{\partial\tau_k}
-\frac{\partial F_{j(\kappa)}^i}{\partial\tau_k}\,x_{j(\kappa)}^i\nonumber\\
&\quad-\frac{\partial F_{j(\kappa)}^i}{\partial\tau_k}\times
\frac{\left(\frac{\kappa}{p}-F_{j(\kappa)}^i\right)^2}{F_{j(\kappa)+1}^i-F_{j(\kappa)}^i}
\,\left(x_{j(\kappa)+1}^i-x_{j(\kappa)}^i\right)\nonumber\\
&\quad+\frac{\left(\frac{\kappa}{p}-F_{j(\kappa)}^i\right)^2}{2\left(F_{j(\kappa)+1}^i-F_{j(\kappa)}^i\right)}
\,\left(\frac{\partial x_{j(\kappa)+1}^i}{\partial\tau_k}
-\frac{\partial x_{j(\kappa)}^i}{\partial\tau_k}\right)\nonumber\\
&\quad-\left(\frac{\partial F_{j(\kappa)+1}^i}{\partial\tau_k}-\frac{\partial F_{j(\kappa)}^i}{\partial\tau_k}\right)
\frac{\left(\frac{\kappa}{p}-F_{j(\kappa)}^i\right)^2}{2\left(F_{j(\kappa)+1}^i-F_{j(\kappa)}^i\right)^2}
\,\left(x_{j(\kappa)+1}^i-x_{j(\kappa)}^i\right).\label{eq:dintxdF2}
\end{align}
We obtain the partial derivative of $X^i$ with respect to $\tau_k$ evaluated at $\kappa/p$ from equations (\ref{eq:dintxdF})--(\ref{eq:dintxdF2}) in the following way:
$$\frac{\partial X^i}{\partial\tau_k}\left(\frac{\kappa}{p}\right)
=\sum_{l=0}^{j(\kappa)-1}\left[\frac{\partial X^i}{\partial\tau_k}(F_{l+1}^i)
-\frac{\partial X^i}{\partial\tau_k}(F_l^i)\right]
+\left[\frac{\partial X^i}{\partial\tau_k}\left(\frac{\kappa}{p}\right)-\frac{\partial X^i}{\partial\tau_k}\left(F_{j(\kappa)}^i\right)\right]~,$$
which enables us to compute the partial derivatives of the sample eigenvalues that belong to the $i^{th}$ support interval with respect to the population eigenvalues as:
\be
\forall \kappa\in\{pF_0^i+1,pF_0^i+2,\ldots,pF_{\omega_i+1}^i\}\quad
\forall k=1,\ldots,p\quad
\frac{\partial\lambda_\kappa}{\partial\tau_k}
=\frac{\partial X^i}{\partial\tau_k}\left(\frac{\kappa}{p}\right)
-\frac{\partial X^i}{\partial\tau_k}\left(\frac{\kappa-1}{p}\right).\label{eq:intgrd}
\ee
This derivation concludes the description of the numerical implementation of the QuEST \mbox{function} and its analytical Jacobian.

\section{Monte Carlo Simulations}
\label{sec:mc}

Section 5.1.1 of \cite{ledoit:wolf:2015} already provides some preliminary evidence documenting the accuracy of the estimator of the population eigenvalues obtained by numerically inverting the QuEST function. The simulations presented below are more extensive. They highlight the convergence rate in log-log scale for various shapes of the population spectrum. 

\subsection{Population Spectrum}
\label{sub:popdist}

The population eigenvalues are taken from the distribution of $1+(\kappa-1)X$, where $\kappa$ is the condition number and $X$ is a random variable whose support is the compact interval $[0,1]$. Throughout the whole simulation study, we carry four different shapes for the distribution of~$X$. 
\begin{enumerate}
\item 
The original shape is left-skewed: it is the \cite{kumaraswamy:1980} distribution with parameters $(3,1/3)$. The Kumaraswamy family is similar in spirit to the Beta family, but more tractable: the density, the c.d.f.\ and the quantile function are all available in closed form. For reference, the c.d.f.\ of Kumaraswamy$(3,1/3)$ is
\be
\forall x\in[0,1]\qquad H_1(x)=1-\left(1-x^3\right)^{1/3}~.
\ee
All the other shapes are derived from this one. 
\item 
The next shape is right-skewed, obtained by taking the mirror image of the density about the midpoint of the support. \citet[p.\ 73]{jones:2009} observes that there is  ``a pleasing symmetry'' in this case: it is equivalent to taking the mirror image of the c.d.f.\ about the 45 degrees line, that is, replacing it with its inverse, the quantile function:
\be
\forall x\in[0,1]\qquad H_2(x)=\left[1-(1-x)^3\right]^{1/3}~.
\ee
\item A symmetric bimodal distribution is generated by combining right-skewness on $[0,1/2]$ with left-skewness on $[1/2,1]$:
\be
\forall x\in[0,1]\qquad H_3(x)=
\begin{cases}
\displaystyle\frac{1}{2}\left[1-\left(1-2x\right)^3\right]^{1/3}&\text{if $x\in[0,1/2]$~,}\\
\displaystyle 1-\frac{\left[1-\left(2x-1\right)^3\right]^{1/3}}{2}&\text{if $x\in[1/2,1]$~.}
\end{cases}
\ee
\item Finally a symmetric unimodal distribution is generated by combining left-skewness on $[0,1/2]$ with right-skewness on $[1/2,1]$:
\be
\forall x\in[0,1]\qquad H_4(x)=
\begin{cases}
\displaystyle
\frac{1-\left[1-(2x)^3\right]^{1/3}}{2}
&\text{if $x\in[0,1/2]$~,}\\
\displaystyle 
\frac{1+\left[1-(2-2x)^3\right]^{1/3}}{2}
&\text{if $x\in[1/2,1]$~.}
\end{cases}
\ee
\end{enumerate}
Note that all four densities diverge to infinity, so the set of shapes chosen is a challenging one.

\subsection{Intuition}

Given the sample eigenvalues $\lambda_{n,1}\leq\lambda_{n,2}\leq\dots\leq\lambda_{n,p}$, we estimate the population eigenvalues $\tau_{n,1}\leq\tau_{n,2}\leq\dots\leq\tau_{n,p}$ by numerically inverting the QuEST function:
\begin{equation} \label{eq:optim}
\widehat{\boldsymbol{\tau}}_n \defeq \argmin_{\mathbf{t}\in[0,\infty)^p}
\,\frac{1}{p}\sum_{i=1}^p\left[q_{n,p}^i(\mathbf{t})-\lambda_{n,i}\right]^2~.
\end{equation} 
The simulation study presented below centers on the base-case scenario where the condition number is $\kappa=10$,  variates are normally distributed, and the concentration ratio is $c=1/3$. For dimension $p=1,000$, Figure \ref{fig:sidebyside} provides a side-by-side comparison of the population spectra specified in Section \ref{sub:popdist} with their sample counterparts.
\begin{center}
\captionsetup{type=figure}
\includegraphics[scale=0.8]{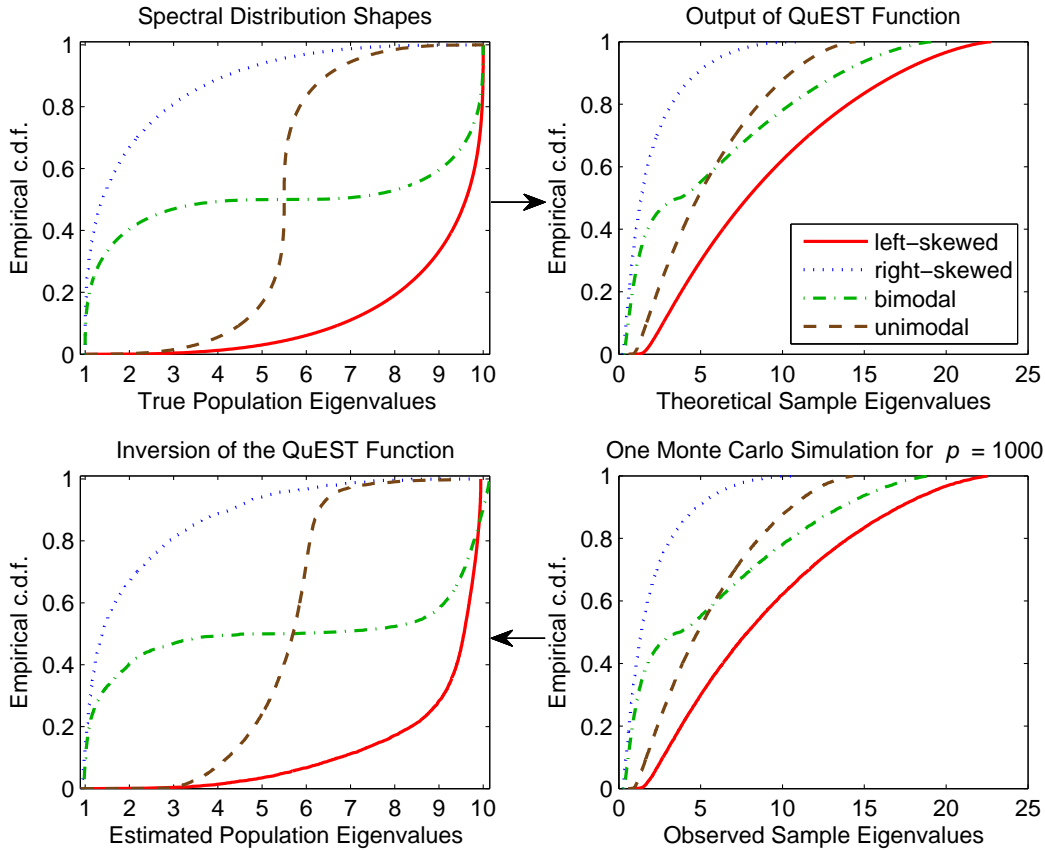}
\captionof{figure}{Population vs.\ sample spectrum. 
  Top panel is the direct QuEST function, bottom panel the inverse of the QuEST function for estimation purposes.}\label{fig:sidebyside}
\end{center}
Let us start with the top panel. It shows the effect of the QuEST function. For each of the four distribution shapes, the population eigenvalues in the top left graph get mapped into the limiting sample spectra shown in the top right graph. This  shows what the Mar\v{c}enko-Pastur transformation does. There is a lot of distortion, but the relative positions of the four color-coded c.d.f.'s have been preserved. Therefore, the information has not been destroyed: it is just waiting to be deciphered by a suitable method. We are essentially facing a severe nonlinear bias-correction problem.

The bottom panel goes in the opposite direction: the QuEST function
gets inverted. At~the bottom right are sample eigenvalues generated in
one Monte Carlo simulation. Observe how closely they match the
nonrandom distributions in the top right. This is because, as
mentioned above, in the large-dimensional asymptotic limit randomness
vanishes. Then numerically inverting the QuEST function yields the
estimator of population eigenvalues shown in the bottom left graph. It
closely matches the truth (shown top left). The distortion has been
undone, and the original shapes of the spectral distributions have
been restored. The bottom panel is our estimation procedure in a
nutshell.

\subsection{Base-Case Scenario}

\citet[Theorem 2.2]{ledoit:wolf:2015} prove that the mean squared
deviation between estimated and true population eigenvalues $p^{-1} \sum_{i=1}^p\left[\widehat{\tau}_{n,i}-\tau_{n,i}\right]^2$
converges almost surely to zero under large-dimensional asymptotics. This quantity is scale-sensitive, whereas the problem is scale-invariant. This is why we study in Monte Carlo simulations the scale-adjusted quantity
\be
\frac{\displaystyle\frac{1}{p}\sum_{i=1}^p\left[\widehat{\tau}_{n,i}-\tau_{n,i}\right]^2}{
\displaystyle\left(\frac{1}{p}\sum_{i=1}^p\tau_{n,i}\right)^2}
\label{eq:consistent}
\ee
instead, called the (empirical) {\it normalized mean squared error}. This change in performance measure does not make any
difference to strong the consistency result, given that
\cite{ledoit:wolf:2015} assume that the population eigenvalues are
bounded away from zero and infinity. But we do not want to give the
visual impression that covariance matrices with a larger trace are
estimated less accurately, since on a relative basis it is not
true. 

The matrix dimension ranges from $p=30$ to $p=1,000$. Convergence of
the scale-adjusted mean squared deviation defined by equation
\eqref{eq:consistent} is displayed in Figure \ref{fig:basecase} on a log-log scale for the four distribution shapes.
\begin{center}
\captionsetup{type=figure}
\includegraphics[scale=\scale]{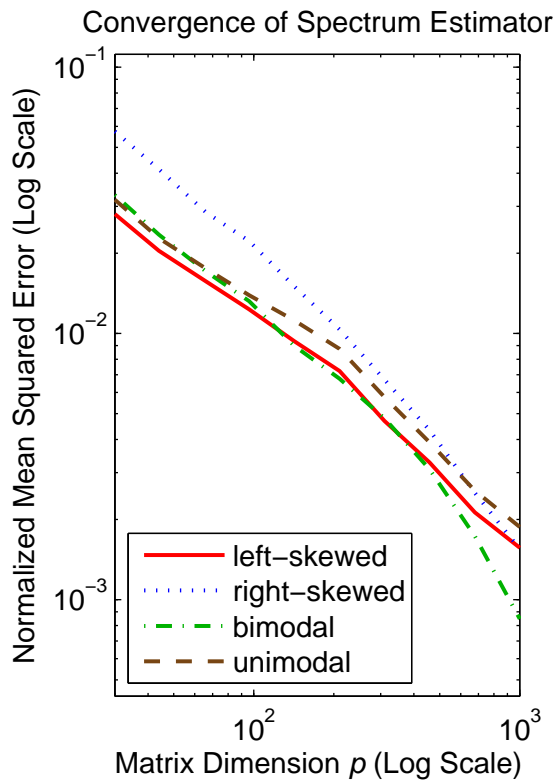}
\captionof{figure}{Consistency of the estimator of population
  eigenvalues in the base case scenario.}
  \label{fig:basecase}
\end{center}
In all log-log graphs presented in this paper, including this one, the scales of the x- and y-axes have been equalized, so that the $-45^\circ$ line corresponds to a convergence rate of $p$.  Each point in the curves corresponds to the average across $1,000$ Monte Carlo simulations. 

In terms of speed, Figure \ref{fig:speed} shows that the numerical
recipe presented in this paper for the implementation of the QuEST
function  is sufficiently fast for practical purposes.\footnote{These
  numbers were run using Matlab R2014b on an Apple Mac Pro with
  a 3.5 GHz Intel Xeon E5 processor.}

\begin{center}
\captionsetup{type=figure}
\includegraphics[scale=\scale]{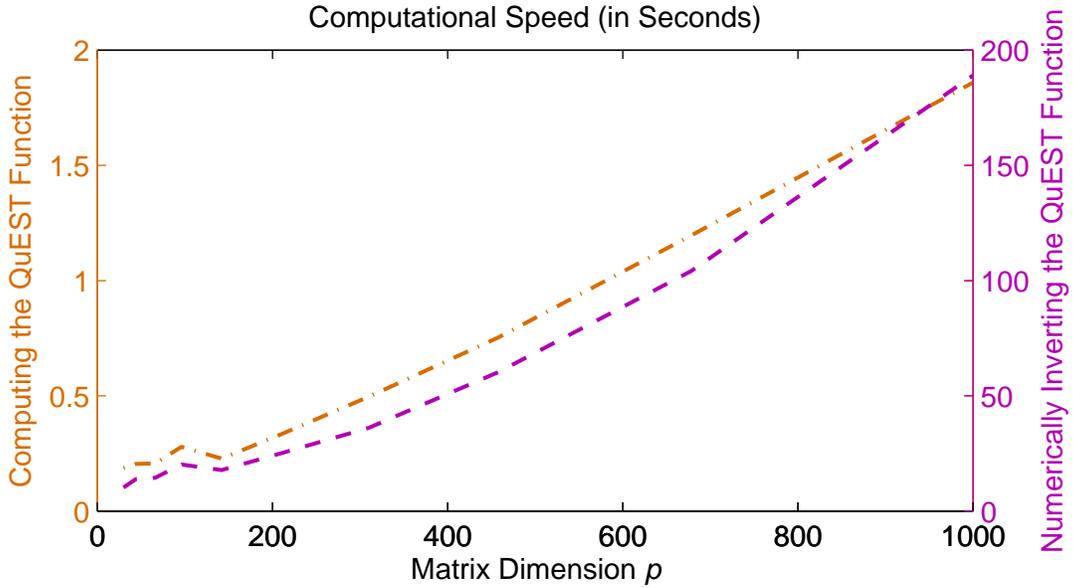}
\captionof{figure}{Speed benchmark for computing the QuEST function and estimating population eigenvalues.}\label{fig:speed}
\end{center}
The remainder of Section \ref{sec:mc} is dedicated to demonstrating
the robustness of the base--case convergence pattern in three
directions: different concentration ratios $c=p/n$, condition numbers
$\kappa$, and variate distributions $D$. 

\newpage
\subsection{Concentration Ratio}

First, we increase the concentration ratio $c=p/n$. We pick two
values: $c=1$ and $c=2$. The first case is not covered by the
mathematical theory of \cite{ledoit:wolf:2015}, but the numerical
results displayed on the left panel of Figure \ref{fig:concentration}
seem to indicate that satisfactory convergence is achieved
nonetheless. In the second case, we manage to consistently estimate
$p$ eigenvalues, in spite of the fact that the sample covariance
matrix has only $n=p/2$ nontrivial eigenvalues. Note that this is the
only graph where we let $n$ (instead of $p$) range from $30$
to~$1,000$, because of $n<p$. 
\begin{center}
\captionsetup{type=figure}
\includegraphics[scale=\scale]{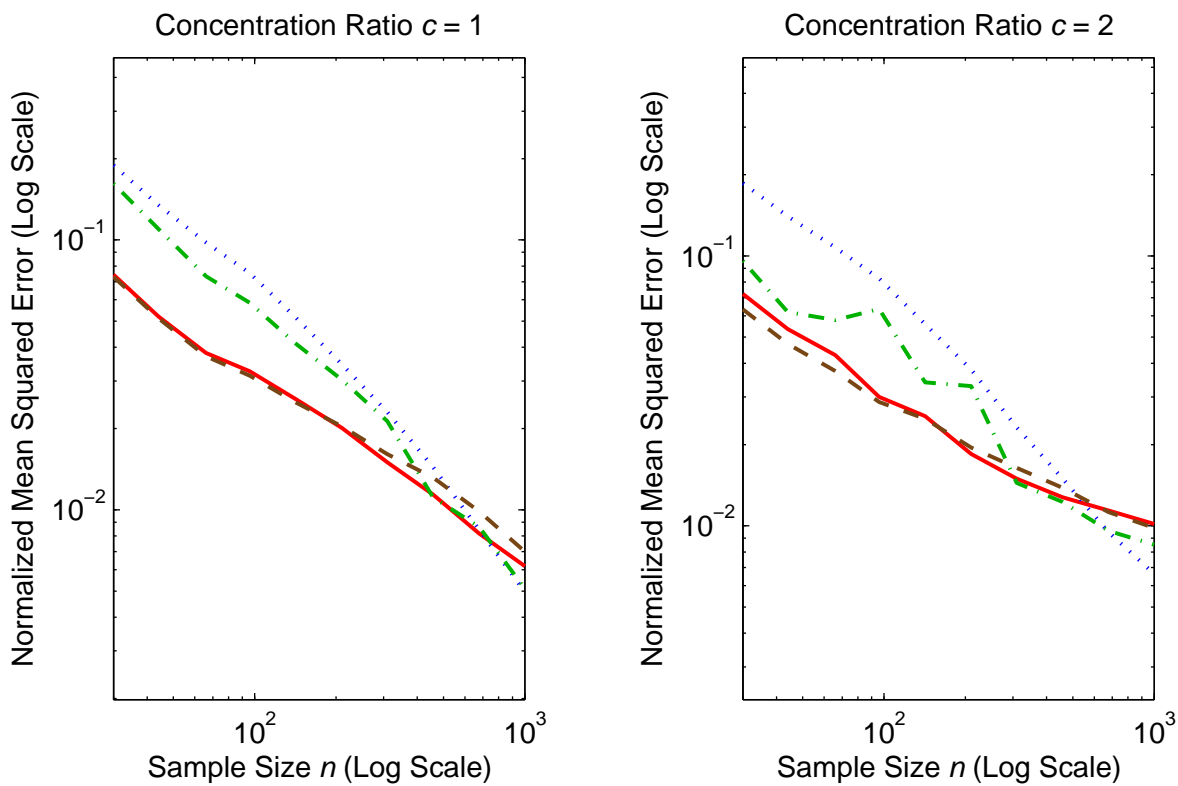}
\captionof{figure}{Consistency of the estimator of population
  eigenvalues for higher concentration ratios.
 Color and line-style code as in Figures~\ref{fig:sidebyside} and \ref{fig:basecase}.}
\label{fig:concentration}
\end{center}

\newpage
\subsection{Condition Number}

The second axis of deviation from the baseline case is to look at condition numbers other than $\kappa=10$. We consider a smaller condition number, $\kappa=2$, and a larger one, $\kappa=100$. The results are displayed in Figure \ref{fig:condition}.
\begin{center}
\captionsetup{type=figure}
\includegraphics[scale=\scale]{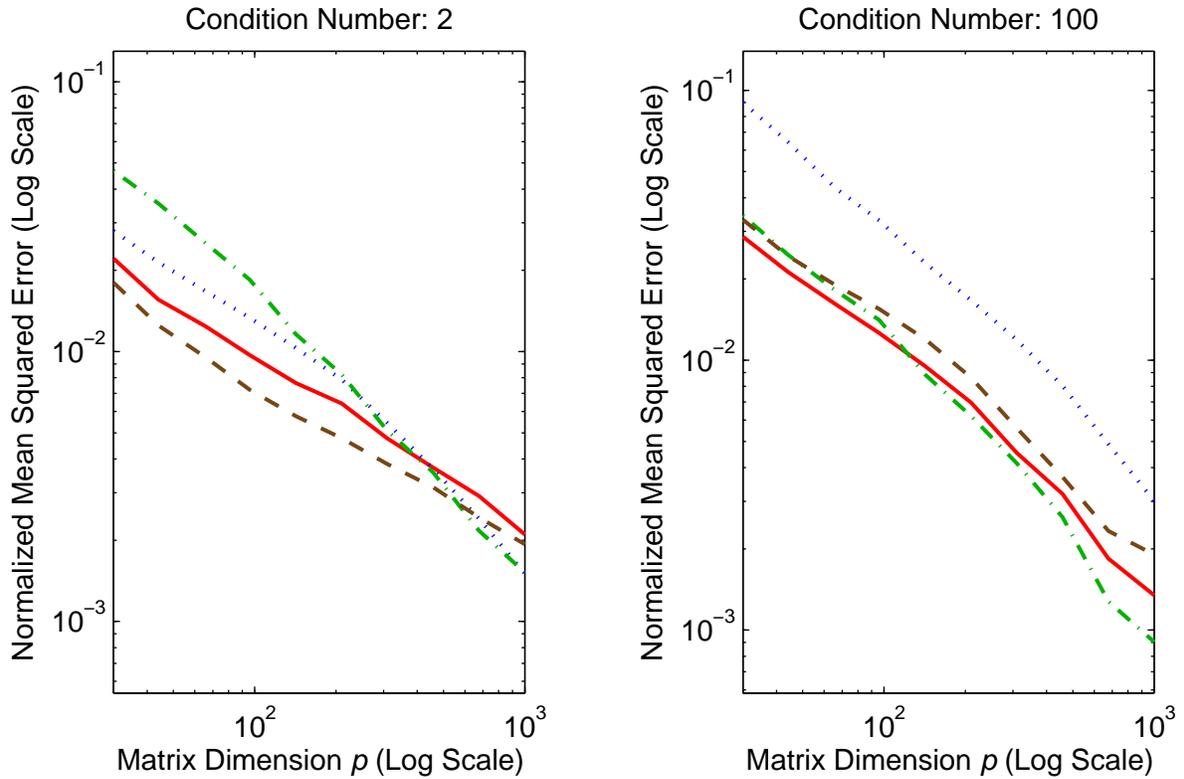}
\captionof{figure}{Consistency of the estimator of population
  eigenvalues for various condition numbers. 
  Color and line-style code as in Figures~\ref{fig:sidebyside} and \ref{fig:basecase}.}
\label{fig:condition}
\end{center}
These results show that we can still obtain convergence in spite of changes in the condition number.

\subsection{Distribution of the Variates}

Finally, we deviate from the base-case scenario in the direction of
having other distributions than Gaussian for the random
variates. First, we take a fat-tailed distribution: the ``Student''
$t$-distribution with $5$ degrees of freedom; and second, the most thin-tailed of all distributions: the Bernoulli coin toss distribution with probability $1/2$. 
The results are displayed in \mbox{Figure \ref{fig:thintail}}.
\begin{samepage}
\begin{center}
\captionsetup{type=figure}
\includegraphics[scale=\scale]{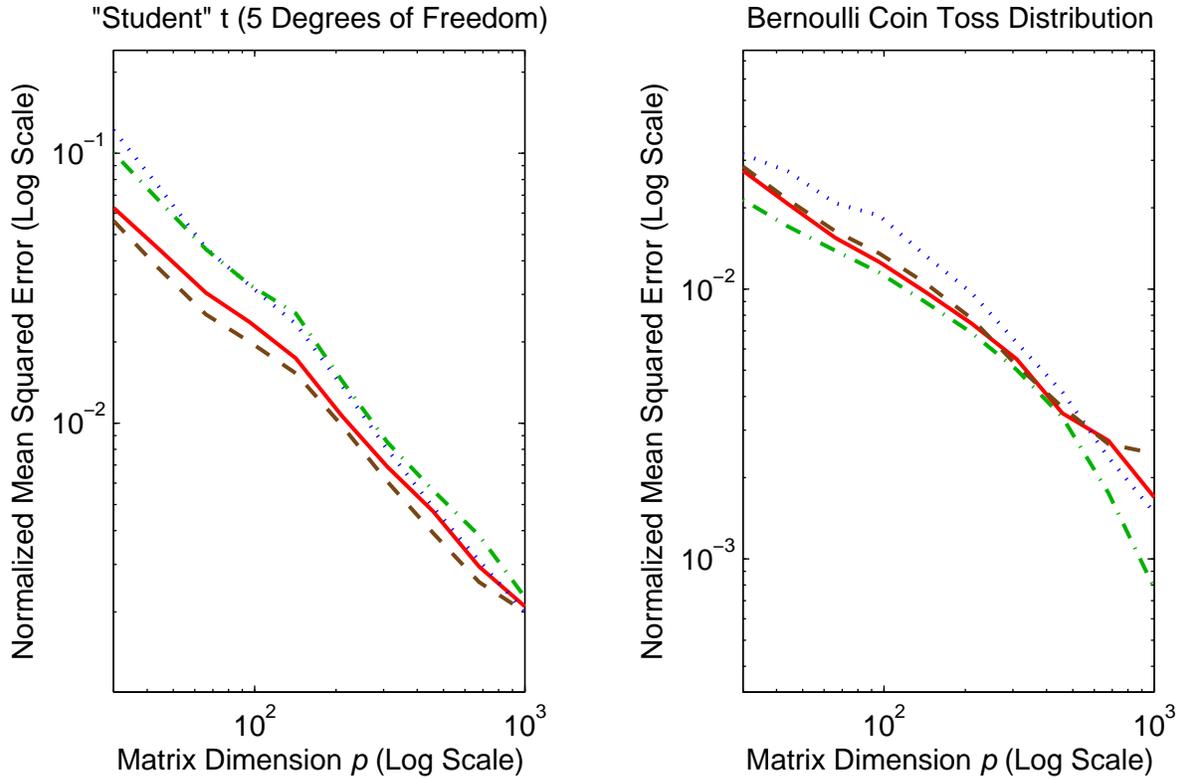}
\captionof{figure}{Consistency of the estimator of population
  eigenvalues when the variates have thick or thin tails.
   Color and line-style code as in Figures~\ref{fig:sidebyside} and and \ref{fig:basecase}.}
\label{fig:thintail}
\end{center}
We also consider a skewed distribution: the exponential. The results
are displayed in Figure~\ref{fig:skewed}.
\begin{center}
\captionsetup{type=figure}
\includegraphics[scale=\scale]{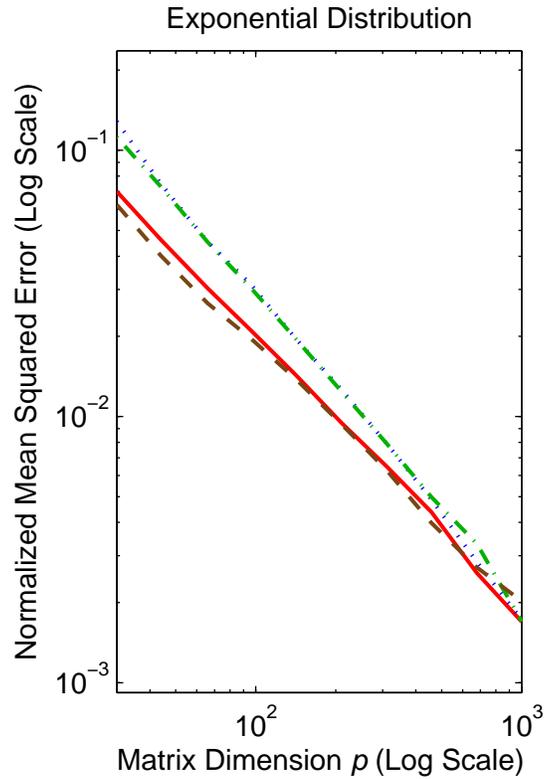}
\captionof{figure}{Consistency of the estimator of population
  eigenvalues when the variates are skewed. 
   Color and line-style code as in Figures~\ref{fig:sidebyside} and and \ref{fig:basecase}.}
\label{fig:skewed}
\end{center}
\end{samepage}
These results show that we can obtain convergence across a variety of
variate distributions.

\subsection{Overview of the Simulation Results}

The Monte Carlo simulations presented above illustrate the ability of the estimator of population eigenvalues constructed by numerically inverting the QuEST function to get closer to the truth as the matrix dimension and the sample size go to infinity together. This exercise has been extensive, involving a grand total of $320,000$ Monte Carlo simulations. The point was to build practical comfort around the theoretical result. Best-fit lines in log-log space have slopes that vary in the range from $-0.70$ to $-1.10$, giving some empirical indication about the exponent of the convergence rate of the mean squared deviation between true and estimated population eigenvalues.

\newpage

\section{Conclusion}
\label{sec:conclusion}

When matrix dimension is not negligible with respect to sample size,
finite-dimension asymptotic approximations are no longer close to the
truth: We enter the \cite{marcenko:pastur:1967} zone instead. In this zone, the sample eigenvalues are a very distorted version of their population counterparts. Only after the publication of \cite{karoui:2008} and \cite{mestre:2008b} did researchers in the field of large-dimensional multivariate statistics start to harbor any hope of unwinding this distortion. 

\cite{ledoit:wolf:2015} put forward a natural discretization of the
Mar\v{c}enko-Pastur \mbox{equation} that can be inverted numerically. Even
though the sample eigenvalues are far from their \mbox{population}
counterparts, the distortion can be inverted through this particular
procedure. The~present paper describes in great detail how to
discretize the Mar\v{c}enko-Pastur equation. We~also provide extensive
Monte Carlo simulations demonstrating the practical effectiveness of
the method in terms of recovering the population eigenvalues. There
are many applications in the field of multivariate statistics,
starting with nonlinear shrinkage estimation of covariance matrices.

\newpage
\bibliographystyle{apalike}
\bibliography{../wolf} 

\end{document}